\documentclass[twocolumn]{aastex631}
\usepackage{xcolor}
\usepackage[T1]{fontenc}

\newcommand\sosay[1]{``#1''}

\newcommand{\rg}{r_{\mathrm{g}}}

\begin{document}

\title{Modulation of X-ray flux by obscuration of neutron star boundary layer}

\author[0000-0003-3958-9441]{G.\,T\"or\"ok}
\affiliation
{Research Center for Computational Physics and Data Processing,\\ Institute of Physics, Silesian University in Opava, Bezru\v{c}ovo n\'am.\ 13,\\ CZ-746-01 Opava, Czech Republic}

\author[0000-0002-0930-0961]{K.\ Klimovi\v{c}ov\'{a}}
\affiliation{Research Center for Computational Physics and Data Processing,\\ Institute of Physics, Silesian University in Opava, Bezru\v{c}ovo n\'am.\ 13,\\ CZ-746-01 Opava, Czech Republic}

\author[0000-0003-0826-9787]{D.\,Lan\v{c}ov\'a}
\affiliation{Research Center for Computational Physics and Data Processing,\\ Institute of Physics, Silesian University in Opava, Bezru\v{c}ovo n\'am.\ 13,\\ CZ-746-01 Opava, Czech Republic}
\affiliation{Nicolaus Copernicus Astronomical Center of the Polish Academy of Sciences,\\ Bartycka 18, 00-716
Warsaw, Poland}

\author[0000-0002-4193-653X]{M.\ Matuszkov\'{a}}
\affiliation{Research Center for Computational Physics and Data Processing,\\ Institute of Physics, Silesian University in Opava, Bezru\v{c}ovo n\'am.\ 13,\\ CZ-746-01 Opava, Czech Republic}

\author[0000-0002-0733-5052]{E.\ \v{S}r\'{a}mkov\'{a}}
\affiliation{Research Center for Computational Physics and Data Processing,\\ Institute of Physics, Silesian University in Opava, Bezru\v{c}ovo n\'am.\ 13,\\ CZ-746-01 Opava, Czech Republic}

\author[0000-0001-9635-5495]{M.\ Urbanec}
\affiliation{Research Center for Computational Physics and Data Processing,\\ Institute of Physics, Silesian University in Opava, Bezru\v{c}ovo n\'am.\ 13,\\ CZ-746-01 Opava, Czech Republic}

\author[0000-0002-3434-3621]{M. \v{C}emelji\'{c}}
\affiliation{Nicolaus Copernicus Superior School, College of Astronomy and Natural Sciences, Gregorkiewicza 3,
87-100, Toru\'{n}, Poland}
\affiliation{Research Center for Computational Physics and Data Processing,\\ Institute of Physics, Silesian University in Opava, Bezru\v{c}ovo n\'am.\ 13,\\ CZ-746-01 Opava, Czech Republic}
\affiliation{Nicolaus Copernicus Astronomical Center of the Polish Academy of Sciences,\\ Bartycka 18, 00-716
Warsaw, Poland}
\affiliation{Academia Sinica, Institute of Astronomy and Astrophysics, P.O. Box 23-141,
Taipei 106, Taiwan}

\author[0000-0002-5085-9740]{R.\ \v{S}pr\v{n}a}
\affiliation{Research Center for Computational Physics and Data Processing,\\ Institute of Physics, Silesian University in Opava, Bezru\v{c}ovo n\'am.\ 13,\\ CZ-746-01 Opava, Czech Republic}

\author[0000-0002-5760-0459]{V.\,Karas}
\affiliation{Astronomical Institute of the Czech Academy of Sciences\\ Bo\v{c}n\'{\i} II 1401, CZ-14100, Prague, Czech Republic}

\begin{abstract}
The quasi-periodic oscillations (QPOs) observed in the X-ray variability of both black hole (BH) and neutron star (NS) systems provide a tool for probing strong gravity and dense matter equations of state. Nevertheless, the mechanism of QPO modulation in NS systems, where the amplitudes of QPOs with frequencies approaching kHz range are very high in comparison to BH high-frequency QPOs, remains an unsolved puzzle. Relativistic ray tracing of photons emitted from the immediate vicinity of compact objects has, to date, been used to investigate various mechanisms that explain the observed weak BH QPOs. However, it has not been applied to model the NS QPO signal, which requires incorporating the NS surface and a bright boundary layer (BL) on it. Here, we explore the QPO modulation mechanisms based on the BL obscuration. Using simplified models of axisymmetric oscillations of thick accretion discs (tori), we demonstrate that the disc oscillations drive the high NS QPO amplitudes through BL obscuration, which is relevant especially for vertical oscillations.  We also demonstrate that obscuration effects enable the observability of the Keplerian frequency in the case of discs that decay due to instabilities.

\end{abstract}

\keywords{stars: black holes --- stars: neutron --- X-rays: binaries}

\section{Introduction} \label{sec:intro}

Accreting black holes (BHs) and neutron stars (NSs) exhibit extraordinarily complex phenomenology of X-ray variability patterns on various timescales, including, among those on longer timescales represented by the low-frequency quasi-periodic oscillations (LF QPOs), the shortest ones of millisecond periods \citep[and references therein]{1998AdSpR..22..925V,2006csxs.book..157M,2016AN....337..398M}, which are usually attributed to the orbital motion. The QPOs with the highest frequencies observed in individual BH sources are commonly denoted as high-frequency (HF) QPOs, while in the case of NS sources, these are often denoted as kHz QPOs, the convention arising from their frequency range of $200$--$1400$ Hz, or twin-peak QPOs, the David Lynch TV series-like name following from the typical profile of the two simultaneous peaks imprinted in power-density spectra (PDS) \citep[][]{2006csxs.book...39V,bel-etal:2012}.

Stellar mass BHs exhibit relatively lower fluxes and variability than NS sources, and the HF QPOs observed in Galactic BH microquasars (hereafter, shortly BH QPOs) occur with root-mean-squared (rms) amplitudes usually not exceeding $5\%$. On the other hand, the rms amplitudes of NS twin-peak QPOs (hereafter, shortly NS QPOs) are often higher and reach maximal values around 30$\%$. The coherence times of NS QPOs are also typically much longer than those of BH QPOs. Based on correlated spectral and timing analysis, it was suggested that these stable features with high signal-to-noise ratio are connected to the periodic changes of the radiation originating in NS boundary layer \citep[BL,][]{2003A&A...410..217G,2005AN....326..812G}.

It is often presumed that QPOs are connected to the orbital motion of hot matter around the central compact object. Observations across a wide range of central compact object masses, from low-mass X-ray binaries to galactic centres, demonstrated that the observed HF QPO frequencies are inversely proportional to the mass of the central object and fall within the range defined by the Keplerian frequency assigned to the innermost parts of accretion discs located in the strong gravity region \citep[][]{remillard_x-ray_2006,2004ApJ...609L..63A,Gol-etal:2019,2023CoSka..53d.175K}. The relativistic effects are crucial for the generation of the QPO signal emerging from the immediate vicinity of the central compact object \citep{1999PhRvL..82...17S,2019NewAR..8501524I}. Consequently, the NS QPOs, along with the correlated low-frequency variability, can be employed to probe strong gravity and put constraints on the dense matter equation of state \citep[EoS,][]{2007PhR...442..109L}. Since most of the NS QPO sources do not show any signatures of NS spin in their variability except the X-ray burst events, it is usually expected that these sources have relatively weak magnetic fields allowing equatorial accretion \citep[see, e.g.,][]{2006csxs.book...39V,2014PhyU...57..377G}. This further supports the viability of the proposed idea of a gravity probe and constraints on the EoS.

Various concepts involving orbital motion were proposed to explain QPOs \citep[e.g.,][]{1999PhR...311..259W, 2001PASJ...53....1K, 1998tx19.confE.315S,2008A&A...487..527C,2001A&A...374L..19A,2003MNRAS.344..978R,2003MNRAS.344L..37R,2004ApJ...617L..45B,2006CQGra..23.1689A,2006MNRAS.369.1235B, 2010MNRAS.405.2447I,2018MNRAS.474.3967D,2017MNRAS.467.4036M}. Here, we briefly recall some of them that are directly relevant to our investigation. The earliest QPO models involve the motion of clumps or blobs of matter in the innermost region of accretion discs, which can modulate the observed light curves as they orbit, having higher temperatures than the surrounding disc. This concept was considered a~potential method to measure the NS mass \citep{1990ApJ...358..538K} and investigated in several works  \citep[][]{1996ApJ...470..743K,1999ApJ...526..953K,2003nlgd.conf..282K}. A specific model based on the kinematics of orbital motion, the relativistic precession model, assigns BH and NS QPOs to the test particle geodesic motion \citep{1999PhRvL..82...17S,1999ApJ...524L..63S}. 

In contrast to the oscillations of individual accreted blobs propagating within the disc, predicated upon the notion of collective oscillations of the accreted material, a family of discoseismic models was formulated \citep{1991ApJ...378..656N,2004PASJ...56..905K}. Apart from these, a substantial variety of rival QPO models was proposed \citep[see, e.g.,][]{2019NewAR..8501524I,tor-etal:2022}. Several of them are related to oscillations of accretion tori. The accretion tori were considered in the context of models assuming their radial and vertical epicyclic oscillations \citep[][] {AbramowiczKluzniak2001,2005A&A...436....1T}, or other oscillation modes \citep{2003MNRAS.344L..37R,2003MNRAS.344..978R,2004MNRAS.354.1040M,blaes_epicyclic_2007,2016MNRAS.456.3245M}.

Recently, a possible connection between the radial oscillations of cusp tori and NS QPOs was explored by \cite{tor-etal:2022}, who noticed that the observed QPO frequencies are well matched when identified with radial precession frequency of a critical cusp torus configuration. They also associate the Keplerian orbital frequency with the possible presence of torus instabilities. In this context, they suggested that QPOs originate in the accreting system evolution, including epochs of stable oscillations of torus along with periods where instability emerges, leading to the formation of an orbiting torus fragment. 



Although many QPO models have been proposed, even the QPO modulation mechanism relevant to NS QPOs and their high amplitudes remains an unresolved puzzle. The concept of modulating the observable NS flux through BL obscuration, which we explore here, may play a crucial role within the framework of several QPO models proposed to date.
Specifically, we investigate the obscuration effects caused by radial and vertical oscillations of thick accretion discs (inner accretion tori) and by orbiting disc fragments. Our findings indicate that the obscuration can explain the high amplitudes of NS QPOs.

\section{Relativistic ray-tracing of boundary layer obscuration}

To achieve a realistic representation of light propagation close to a compact object, it is necessary to numerically solve the null geodesic equations and simulate the paths of photons in curved spacetimes using a~relativistic ray-tracing code 
\citep{1973ApJ...183..237C,2005MNRAS.359.1217B,2006ApJ...637L.113S,2011CQGra..28v5011V,2009ApJ...696.1616D,2023ApJ...950...35P,1992MNRAS.259..569K}.

Within the context of the QPO modelling, ray-tracing was pioneered by \cite{2004ApJ...606.1098S,2004ApJ...617L..45B,2005ApJ...621..940S, 2005AN....326..849B,2006ApJ...642..420S,2006ApJ...651.1031S,2014MNRAS.439.1933B,Bakala+2015b,2017MNRAS.467.4036M}. However, these studies were relevant to the case of weak BH QPOs while including a~BH within the simulated system and not including the presence of a~NS with its luminous BL.

In this work, we build on these previous results but include an NS with bright BL, focusing on its obscuration by an optically thick accretion tori, and explore the main differences between the NS and BH case. We are using the \texttt{LSD} \citep[Lensing Simulation Device,~][]{Bakala+2015b} ray-tracing code. For simplicity, we adopt the Schwarzschild geometry describing non-rotating NSs and BHs.

\section{Overall geometry, star and disc}

We assume such a radially transient geometry of the accretion system \citep[][]{1995ApJ...445L..43M, 2006A&A...447..813F,2016MNRAS.461.1967I}, where the extended outer thin disk does not reach the innermost stable circular orbit (ISCO), but transitions into a geometrically thick torus in the innermost region. The overall geometry of the accreting system is depicted within the schematic Figure~\ref{figure:system}. Next, we aim to compare a system with a central NS with a bright BL on its equator, an inner thick torus, and an extended thin accretion disc to a similar system but with central BH and no NS surface with BL.

{For NS systems it is commonly expected that more than $60\%$ of the total emitted accretion energy is released below the disc as the accreted fluid approaches the NS surface via BL, $P_\mathrm{BL}\gtrsim0.6P_\mathrm{total}$  {\citep{1986SvAL...12..117S}}.} This expectation and the integration of the thin-disc flux equation \citep{1973A&A....24..337S}, along with its significant radial extension, determines the total luminosity of the system. To determine the variability of the observable signal, we model the behaviour of the central parts of the system, as well as the behaviour of the outer disc. 

\section{Neutron star and boundary layer}

A spherical non-rotating compact star is considered, with radius $R_\mathrm{NS} = 4.8\,\rg$.\footnote{We measure distances in the units of gravitational radii, $\rg=GM/{c^2}$, where $M$ is the mass of the central object, $G$ the gravitational constant and $c$ the speed of light (the event horizon radius of a Schwarzschild BH is located at $r_{\mathrm{G}}=2\rg$).} This radius is well below the ISCO ($r_{\mathrm{ISCO}}=6\,\rg$){, allowing} for moderate radial oscillations of the inner accretion tori. This choice of radius also satisfies various NS EoS \citep{2019ApJ...877...66U,mat2024a}. We assume a low magnetised NS, where the disk's material accretes on the equator of the star and not through accretion columns on the magnetic poles. A luminous equatorial BL is formed in this configuration, where the hot material flows and spreads on the star's surface \citep{1999AstL...25..269I}.

\begin{figure*}
    \centering
    \includegraphics[width=1\linewidth]{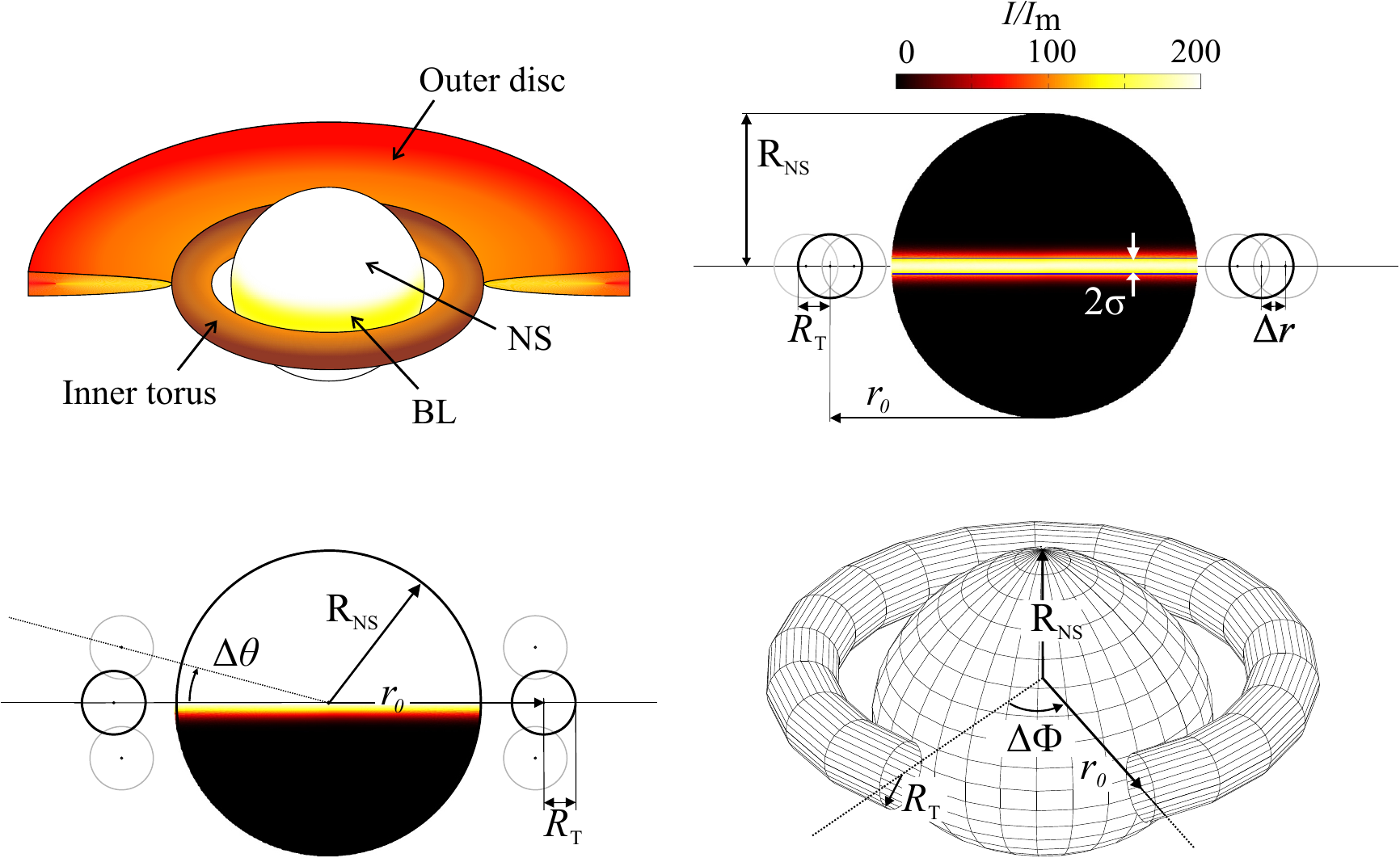}
    
    \caption{Schematics of the considered setup. \textit{Top left:} An overall illustration showing components of the investigated system of accreting NS. \textit{Top right:} The NS BL and radial oscillations of the torus. \textit{Bottom left:} The vertical oscillations of the torus.  \textit{Bottom right:} The orbiting fragment of torus characterised by its opening angle $\Delta\Phi$ (BL on NS surface is not shown).
    \label{figure:system}}
\end{figure*}

The four-velocity of the material in the BL at the equator corresponds to the Keplerian value of the specific angular momentum at ISCO. As the material approaches the poles, it converges towards the four-velocity of the star's surface, with the same profile as the local emissive power (total flux from {unit} surface area), $I=\mathrm{d}\phi/\mathrm{d}S$. In our simplified case of a non-rotating star, the material at the pole is completely decelerated. The local emissive power, expressed in terms of the radial maximum of thin disc emissive power $I_\mathrm{m}$ \citep{1986SvAL...12..117S}, is chosen to peak at the value of  $I=190\,I_\mathrm{m}$  at the equator. It decreases towards the poles, where it vanishes, which is approximated by Gaussian distribution function.{\footnote{{This along with the other setup described within the next sections ensures that $P_\mathrm{BL}\approx0.6P_\mathrm{total}$.}}} Although more sophisticated spreading BL models can be considered, this simplified setup is sufficient for our study \citep[e.g.,][]{2006MNRAS.369.2036S,2014PhyU...57..377G}. The BL emissive power distribution  on the NS is illustrated in the top right panel of Figure~\ref{figure:system}.

\section{Black hole}

In the case of the configurations representing systems including a BH, no surface is located in the centre, only a non-rotating BH. Using relativistic ray tracing, all effects of special and general relativity are included in both NS and BH simulations, such as the Doppler shift, light-bending and higher-order images. In contrast, in the BH case, strong gravity effects are more prominent due to the absence of a star's surface.

\section{Torus and its kinematic motion}
\label{section:kinematics}

Oscillating tori are often considered when modelling the NS systems variability \citep[e.g.][]{2001A&A...374L..19A,2003MNRAS.344L..37R,2003MNRAS.344..978R,2018MNRAS.474.3967D,2016MNRAS.461.1356F}. While a variety of possible torus oscillatory modes exist, including axisymmetric and non-axisymmetric modes and their combinations, we focus on two simplest cases of axisymmetric radial and vertical motion. In addition, we also investigate an orbiting fragment of a torus. These three individual kinematic effects can provide a basis for further studies of obscuration effects within various QPO models.

\subsection{Torus configuration}

For the inner geometrically thick fluid torus, we adopt the methodology introduced in \cite{2004ApJ...617L..45B} and \cite{2005AN....326..849B}. There, tori with a circular cross-section were assumed. In our case, we expect the torus to be effectively optically thick.

Motivated by several studies that related the power-spectral properties of most NS QPOs to phenomena connected to specific radii close to the ISCO \citep[][]{1999PhRvL..82...17S,2005AN....326..808B,2006MNRAS.370.1140B,2009A&A...497..661T,2016MNRAS.457L..19T,tor-etal:2022}, we locate the torus centre, where the flow orbits with Keplerian velocity, at $r_0 = 6.75\,\rg$.{\footnote{{At this radius, the Keplerian, vertical and radial epicyclic frequencies are related as $\nu_\mathrm{K}=\nu_\theta=3\nu_r$. Moreover, the radial epicyclic frequency here rapidly increases with $r$, as it vanishes at $r_\mathrm{ÏSCO}=6\,\rg$ and reaches a maximum at $r=8\,\rg$.}}} Accordingly, we set the torus radius, $R_{\mathrm{T}}$, to $1\,\rg$. In the framework of pressure-supported torus theory, this torus size corresponds to the critical cusp torus at a given radius and thus represents the limiting size \citep[][]{mat2024a}. 

The fluid four-velocity on the surface is given by the condition that the torus fluid's specific angular momentum is constant. For simplicity, for all ray-tracing simulations involving the kinematic motion of the torus, we assume that the torus emissive power integrated over the whole surface (overall luminosity) is constant in time with respect to a distant observer. Furthermore, we assume a lower radiative efficiency of the torus \citep[][]{2016MNRAS.456.3915S} so that the total luminosity corresponds to 10 $\%$ of the power that is radiated by the inner part of a thin disc which is replaced by the torus in its equilibrium position.

\subsection{Radial and vertical oscillations of torus}

Following the concept introduced within QPO models of \cite{KluzniakAbramowicz2001,2004AIPC..714...21A,2004ApJ...617L..45B,Bakala+2015b},  the centre of the torus oscillates in the radial or {vertical} direction with frequencies $\nu_{r}$ and $\nu_{\theta}$ respectively, corresponding to the radial and vertical epicyclic frequencies of free test particles. The amplitudes of oscillations are considered as $\Delta r = 0.75\,\rg$ and $\Delta \theta =15^\circ$. These values were chosen arbitrarily but correspond to oscillation amplitudes that are quite high but still physically conceivable.

The schematics of the oscillating torus and its parameters are shown in the top right and bottom left panels of Figure~\ref{figure:system}. {These schematics are valid for both axisymmetric and non-axisymmetric modes of torus oscillations. 
In our ray-tracing simulations, we consider only the two separate axisymmetric modes.
Thus, at each moment of the oscillatory motion, during radial oscillations, all parts of the torus uniformly either approach or recede from the neutron star, while during vertical oscillations, they move coherently upward or downward.} 

\subsection{Orbiting fragment of torus}

In addition to the oscillating torus, we assume an orbiting fragment of the torus, which is motivated by a scenario in which a~torus instability leads to the disintegration of the torus to fragments, which orbits the star with Keplerian velocity \citep[][] {1984MNRAS.208..721P,1986MNRAS.221..339G,1987MNRAS.225..695G, tor-etal:2022}.  We consider the case with just one torus fragment. This simple consideration directly corresponds to the instabilities related to the torus lowest-order oscillation modes, and its investigation can also inform further research on similar but more complex configurations.  The torus fragment geometry is shown in the lower right panel of Figure~\ref{figure:system}. Since the torus opening angle $\Delta \Phi$ should mostly only affect the harmonic component of the signal, we arbitrarily set $\Delta \Phi=\pi/3$.

\section{Outer disc}

The outer disc in the adopted configuration corresponds to the standard Shakura-Sunyaev thin disc \citep{1973A&A....24..337S} with emissive power $I_{D} = C/r^{3} \left(1 - \sqrt{6{\rg}/r}\right)$, where $C$ is a normalization constant \cite[e.g.,][]{2002apa..book.....F}. The outer edge is located at the radius of $200\,\rg$, which is large enough not to affect the results of the ray-tracing simulations. 

The disc then terminates at the outer edge of the inner torus. This depends on the above-described torus configuration. For the vertical oscillations and torus fragments, it is located at $7.75\,\rg$. In the radial oscillation case, the disc's inner edge moves with the torus, i.e., from $7\,\rg$ to $8.5\,\rg$, so it matches the maximum extent of the inner torus.

\section{Results of ray-tracing simulations}
\label{section:simulations}

The setup described above allows us to focus on the obscuration by the torus, which is close to the ISCO and the surface of the central NS. It assumes a large torus size, but the star is compact, so radial oscillations can have a large amplitude. In such a close star-torus configuration, the observer's inclination angle $i$ is the main parameter affecting the results. In this section, we first examine the light curves in the case of $i=60^\circ$, for which the obscuration effects are well visible but not extremely strong. We then examine the effect of inclination and, finally, the nature of the observable signal.

\subsection{Variability}

\begin{figure}
    \centering
    \includegraphics[width=1.\linewidth]{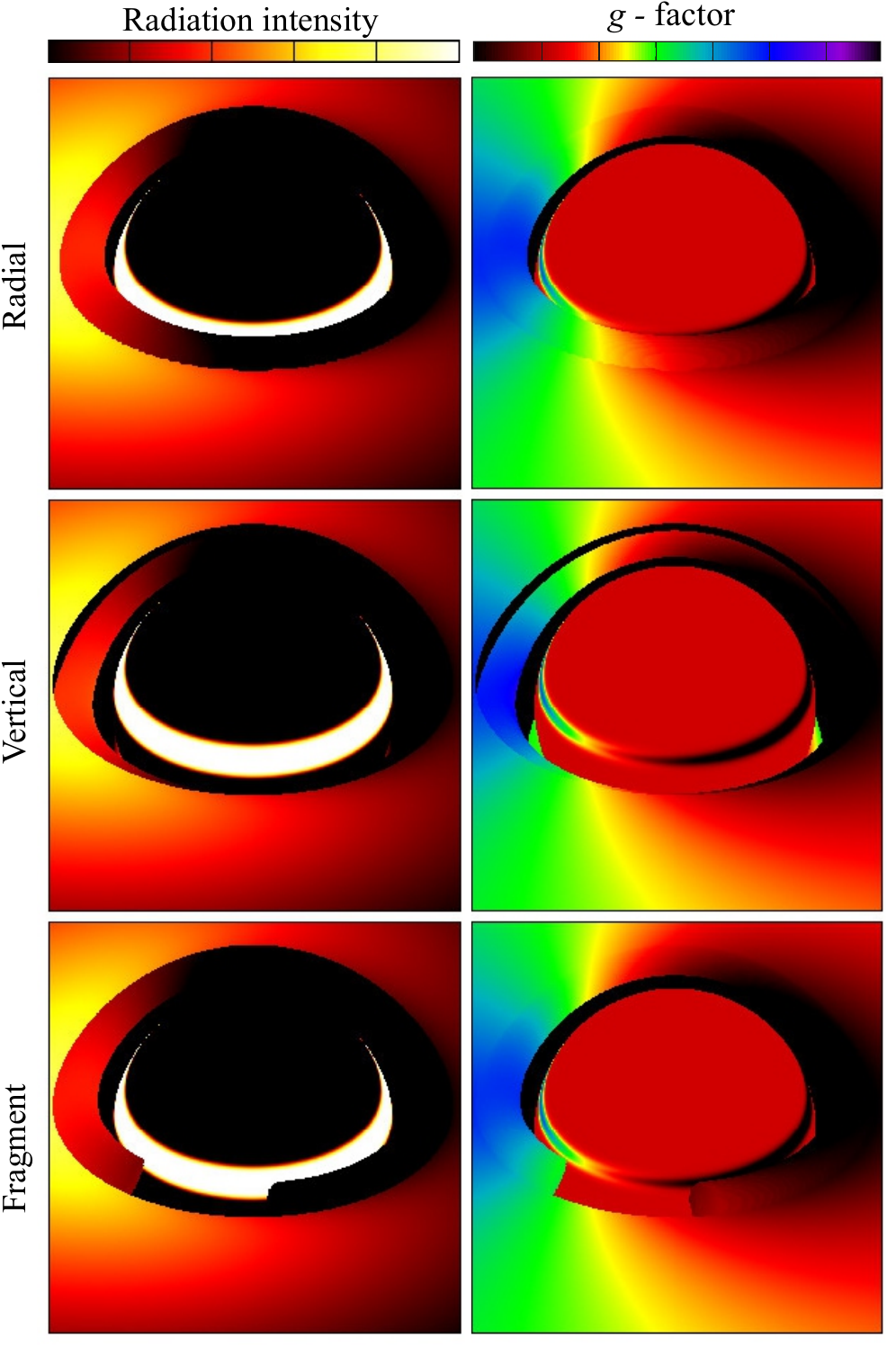}
    \caption{Snapshots from relativistic raytracing simulations showing {the} inner regions of the accretion system of NS as seen by distant observer exploring {the} radially oscillating torus  (top panels), {the} vertically oscillating torus (middle panels), and {the} orbiting fragment of torus (bottom panels) {from the observer's inclination angle $i=60^\circ$}. \textit{Left:} Radiation intensity colour-maps. \textit{Right:} Colour-maps of g-factor.  }
    \label{fig:7:INT}
\end{figure}

\begin{figure*}
    \centering  \includegraphics[width=1\linewidth]{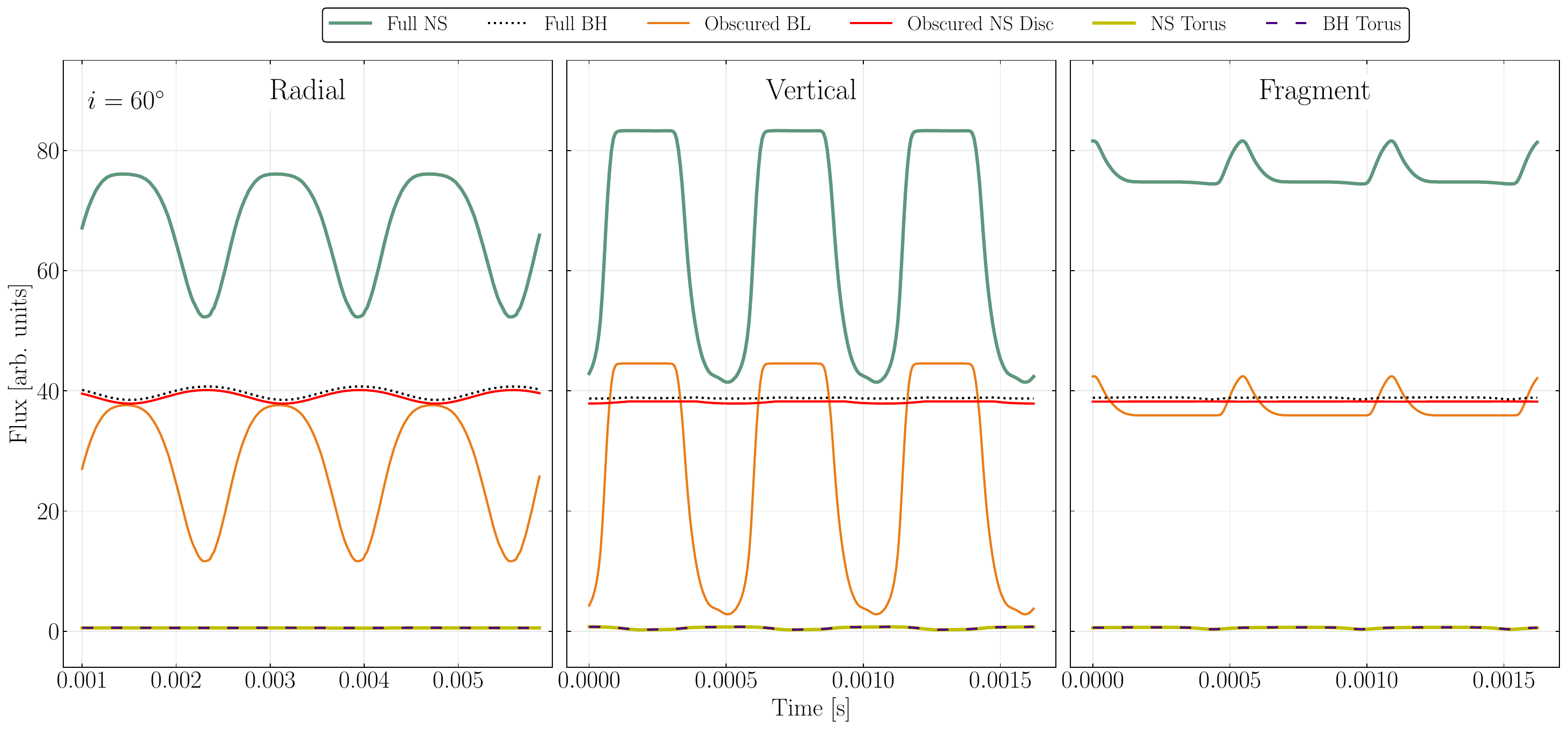}

\caption{Light curves obtained for radial (\textit{left}) and vertical (\textit{middle}) oscillations and for the orbiting torus fragment (\textit{right)} for inclination angle $i=60^\circ$. The extension corresponds to one Keplerian period at $r_0$ for the radial case and three for the vertical and fragment cases. The light curves for individual system components are shown, with the obscuration effects preserved. 
The dark {green} line corresponds to the full model with NS, while the black {dotted} curve shows the BH case. {The yellow curve (NS) and the dashed violet curve (BH) then show only the radiating torus.} The orange one is the NS with BL obscured by non-radiating torus. The red line then shows the outer disc obscured by a non-radiating torus and {NS}.}
    \label{fig:5:lightcurves}
\end{figure*}

Within Figure~\ref{fig:7:INT}, we show snapshots from relativistic raytracing simulations of the considered kinematic effects drawn as colour maps of radiation intensity and frequency shifts (g-factors).  Figure~\ref{fig:5:lightcurves} then presents the resulting variability by light curves. For each of the three investigated kinematic effects, the Figure shows partial light curves comprising the radiation and obscuration of various components of the examined systems. These include the two partial lightcurves accounting only for the torus radiation and all obscuration effects in NS and BH cases (self-obscuration, obscuration by disc, obscuration by NS). Then, we include two light curves obtained when the disc or BL radiation is included. Finally, the two full light curves produced by the complete NS or BH system are shown.

Examining Figure~\ref{fig:5:lightcurves}, we can see that for all three kinematic effects, there is some, but very weak, light curve modulation when the torus is emitting alone. In the BH case, a final variable light curve arises due to the obscuration of the disk by the torus, which is particularly relevant for the modulation given by the radial oscillations of the torus. In the case of NS, the periodic obscuration of the BL by the torus clearly plays the most important role. 

\subsection{Impact of inclination angle}

 \begin{figure}
     \centering
     \includegraphics[width=1\linewidth]{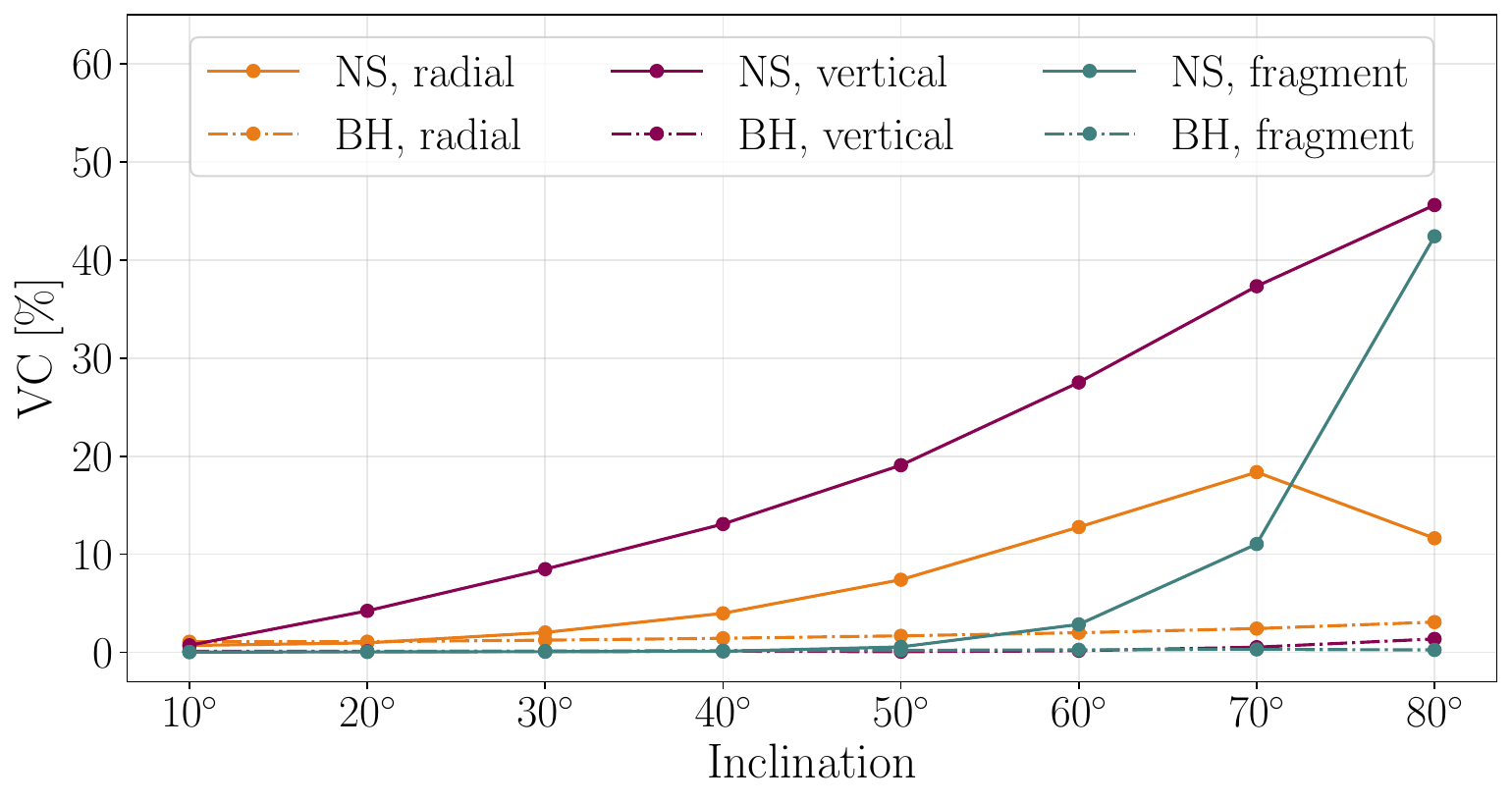}
     \caption{Dependence of the variation coefficient of the light curves on the inclination angle of the observer calculated for the radial and vertical oscillations and orbiting torus fragment. The NS and BH configurations are compared.}
     \label{fig:inclinations}
 \end{figure}

We repeated the above-described analysis for a range of inclinations between the face-on and the edge-on view. We calculated the variation coefficients of the resulting light curves as the ratio of the standard deviation to the mean of flux. The output of this procedure is shown in Figure~\ref{fig:inclinations}. From the Figure, we can see that for the two kinematic effects involving only equatorial motion, namely the radial oscillations of the torus and the Keplerian motion of the torus fragment, the variability of the light curves monotonically increases with $i$. For the vertical oscillations of the torus, the variability is higher since the non-equatorial motion is involved, and the dependence on $i$ has a maximum close to $i=70^\circ$.

 \begin{figure*}
     \centering
     \includegraphics[width=1\linewidth]{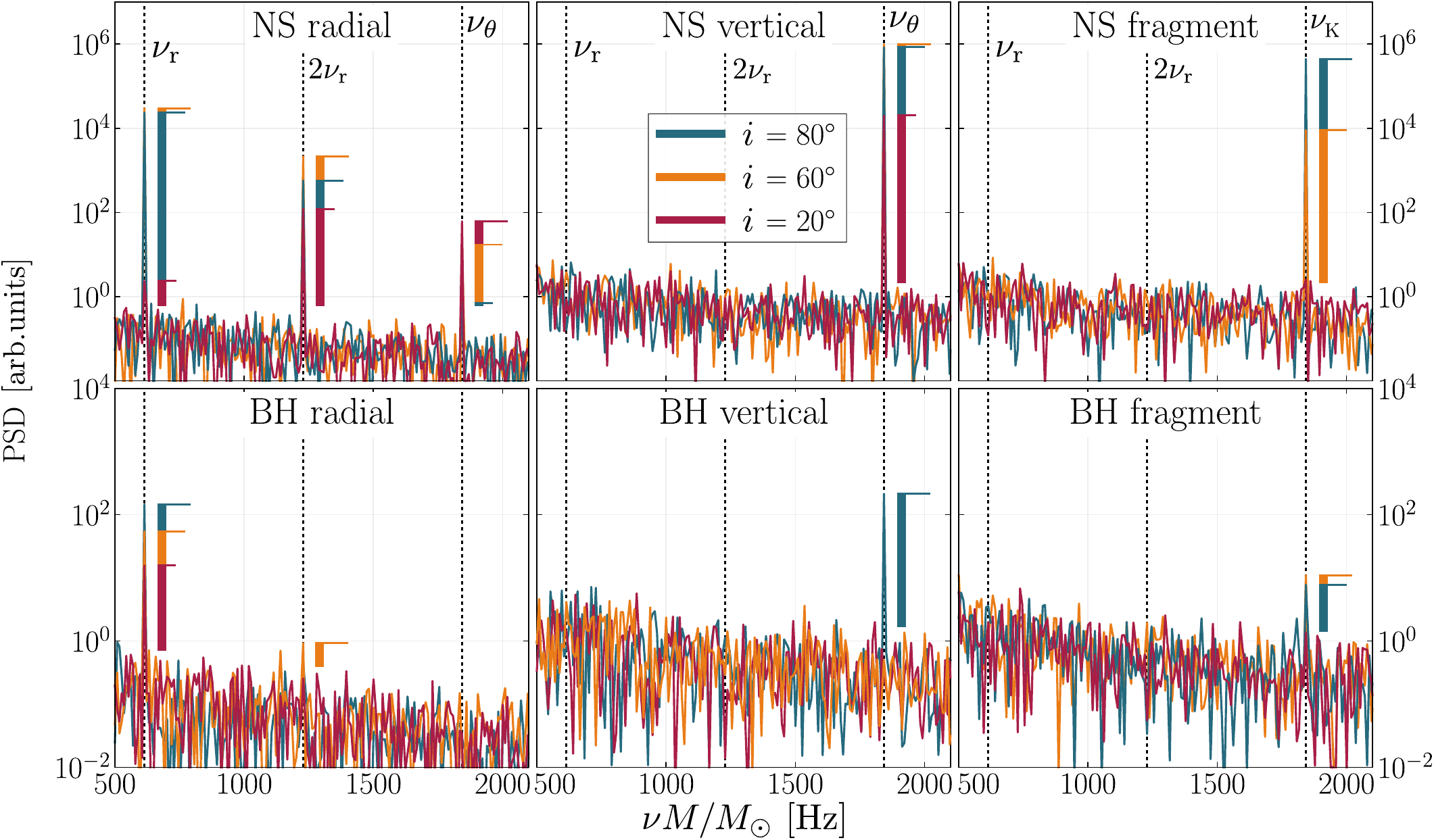}
     \caption{Simulated power density spectra for inclinations $i=80^\circ$ ({teal}), $i=60^\circ$ (orange) and $i=20^\circ$ ({maroon}) for the radial and vertical oscillations and the orbiting torus. The NS and BH cases are compared. The black dashed lines denote the radial epicyclic frequency at the torus centre $\nu_\mathrm{r}$, its harmonics, and the vertical epicyclic frequency $\nu_\theta$, which coincides with the Keplerian frequency $\nu_\mathrm{K}$. For {the} given choice of radius, there is $\nu_\mathrm{K}=\nu_\theta=3\nu_r$. 
     }
     \label{fig:5:psd}
 \end{figure*}

\subsection{Observable signal}
\label{section:PDS}

Up to this point, the results we obtained above in the form of pure model light curves are independent of assumptions about the overall temporal behavior of the whole accreting system. To approximately illustrate their impact on the observable flux, we estimate the predicted observable PDS of our simulations following the approach of \cite{1999ApJ...526..953K}, \cite{2014MNRAS.439.1933B} and \cite{2023CoSka..53d.175K}. The lightcurves obtained from ray tracing simulations are combined with a~typical X-ray background noise associated with the hard spectral state of an X-ray binary source.  The PDS resulting from such total synthetic flux associated with the NS sources  are presented in Figure~\ref{fig:5:psd}. 

\section{Discussion and conclusions}

Within the investigated scenarios, the torus acts as a complex \sosay{clock} that determines the observed variability, while the BL acts as an \sosay{amplifier} enhancing the signal.
Comparing the results obtained for NS and BH cases, our results show that the periodic obscuration of BL allows for higher QPO amplitudes and sets apart the signal produced in an NS source from that produced in BH systems.
Inspecting Figures \ref{fig:inclinations} and \ref{fig:5:psd}, we find that the significant impact of the BL amplification arises for the observer inclinations $i\gtrsim20^\circ$.


\subsection{Implications for particular models of QPOs}

It was suggested that a strong modulation of BL flux could be given by accretion rate modulation by torus oscillations \citep[][]{2005tsra.conf..604K,2007AcA....57....1A}. The accretion rate modulation due to the axisymmetric radial oscillations of cusp tori was confirmed in magnetohydrodynamic simulations of \cite{2017MNRAS.470L..34P}. However, they identified an apparent absence of modulation caused by vertical oscillations. Our results suggest that the obscuration can recover the frequency peaks corresponding to the vertical oscillations. Moreover, as we explored, the obscuration is relevant for both vertical and radial oscillations. Thus, a complex investigation should be conducted to account for the two modulation mechanisms simultaneously. 

Within the framework of models accounting for torus instabilities caused by torus oscillations or other effects, our findings show that the instability product can strongly imprint the Keplerian frequency in the observed light curve. 

\subsection{Modeling the observed variability}

In our work, we focus on three scenarios of BL layer obscuration by single kinematic effects, namely mechanisms of radial and vertical axisymmetric oscillations of a geometrically thick torus in the innermost parts of the accretion disc and the Keplerian motion of a torus fragment.
Combined motion mechanisms, such as simultaneous radial and vertical oscillations or oscillations of the orbiting torus fragment, should be explored in further BL obscuration studies attempting to reproduce the observed PSDs within specific twin-peak QPO models. 
Moreover, it may be useful to include non-axisymmetric oscillatory modes of the torus. In each case, more attention should also be paid to the harmonic content of the signal.

\newpage

\subsection{Summary and overall implications}

 This work investigated the effects of BL obscuration on the observable  X-ray variability of NS X-ray binaries. In addition to the BL obscuration, we included all mutual obscuration effects between the BL, NS, torus and the outer accretion disc. Despite the work on combined motion and non-axisymmetric modes yet to be done, our results show that incorporating the BL obscuration modulation mechanism into a particular QPO model can resolve the issue of high amplitudes of the observed NS QPOs.

This conclusion is relevant for a wide range of models, including the above-mentioned epicyclic models \citep{AbramowiczKluzniak2001,KluzniakAbramowicz2001}, the model of \cite{tor-etal:2016:MNRAS} and \cite{tor-etal:2022}, models by Rezzolla and collaborators \citep{2003MNRAS.344..978R}, or the model explored by Ingram and collaborators \citep{2010MNRAS.405.2447I}.

\begin{acknowledgments}

{GT, KK, DL, MM, ES, MU and RS thank Pavel Bakala for the opportunity to share spacetime, rooms, labs, trains, flights, manuscripts, grants, equations, drinks, comps, ideas and fragments of life. We thank W{\l}odek Klu{\'z}niak, John Miller and Tobias Fischer for discussions. We are grateful to the referee for valuable comments and suggestions, which significantly helped to improve the paper.} We acknowledge the Czech Science Foundation grant{s} no.\ GA\v{C}R~21-06825X {and GA\v{C}R~25-16928O and the Polish NCN grant no.~2019/33/B/ST9/01564}.
We also thank {to the MSK grant no. 04076/2024/RRC} and the internal grants of Silesian University, $\mathrm{SGS/31/2023}$ and $\mathrm{SGS/25/2024}$.

\end{acknowledgments}

\bibliography{tor-etal}{}
\bibliographystyle{aasjournal}

\end{document}